# Synthesis of $RuSr_2(Ln_{3/4}Ce_{1/4})_2Cu_2O_{10}$ magneto-superconducting compounds with Ln = Ho, Y and Dy by high-pressure high-temperature (HPHT) technique

V.P.S. Awana and E. Takayama-Muromachi

Superconducting Materials Center (Namiki Site), National Institute for Materials Science, 1-1 Namiki, Tsukuba, Ibaraki 305-0044, Japan

The samples of composition $RuSr_2(Ln_{3/4}Ce_{1/4})_2Cu_2O_{10}$ with Ln = Ho, Y and Dy. being synthesized by high pressure high temperature (6Gpa, 1200 $^0$C) solid state synthesis route do crystallize in space group *I*4/*mmm*. These samples exhibit magnetic transitions at ~150 K with significant branching of zfc (zero-field-cooled) and fc (field-cooled) magnetization and a sharp cusp in zfc at ~ 100 K, followed by a superconducting transition at lower temperatures. All the compounds show typical ferromagnetic hysteresis loops in magnetic moment ($M$) versus field ($H$) magnetization at 5 K. Near saturation of the moments at 5 K is achieved in above 5 T (Tesla) applied fields with a value of ~ 0.80 $\mu_B$. At low temperatures (5 K) these compounds exhibit both superconductivity and ferromagnetism. To our knowledge these are first successfully synthesized Ru-1222 compounds with various lanthanides including Y, Dy, and Ho. The results are compared with widely reported Gd/Ru-1222 compound.

## 1. INTRODUCTION

The co-existence of magnetism and superconductivity was first reported in $RuSr_2Gd_{1.5}Ce_{0.5}Cu_2O_{10}$ (Gd/Ru-1222) compound [1], and later in $RuSr_2GdCu2O_8$ (Gd/Ru-1212) [2]. Though the phenomenon was discovered later in Gd/Ru-1212, it is this compound which is studied more rigorously than Gd/Ru-1222. Bulk nature of the ferromagnetic order parameter in Gd/Ru-1212 was evidenced from muon-spin-resonance (μSR) and electron-spin-resonance (ESR) experiments [2,3]. Neutron diffraction experiments for Ln/Ru-1212 (Ln = Gd, Eu, Y) indicated that they have a G type antiferromagnetic structure with $\mu_{Ru} \approx 1\mu_B$ and ferromagnetism in these phases is due to canting of the Ru moments [4,5].

As far as the phase purity is concerned, good amount of work is done on Gd/Ru-1212 including synchrotron x-ray [6], neutron powder [7] diffractions, and high resolution transmission electron microscopy [6]. Yet the question of phase purity is not resolved to the satisfactory level [8]. Not only various phase pure non-superconducting samples do exist [9], but also the reproducibility of superconducting compounds with same heat treatments is reported in doubt [10].

To see the magnetic behaviour, Gd/Ru-1212 is not a proper system because of the presence of magnetic Gd ions ($8\mu_B$) which hinders in knowing the exact magnetic contributions form the Ru ions and form superconductivity. Ru-1212 can be formed for non-magnetic Y instead of Gd, but only with high-pressure high-temperature (HPHT) synthesis technique [5,11]. Ironically for Ru-1222 no samples are yet synthesized with HPHT synthesis route. It is our aim here to synthesize and study the magnetic properties of the superconducting Ln/Ru-1222 samples with various Ln including non-magnetic Y. All the samples with composition $RuSr_2(Ln_{3/4}Ce_{1/4})_2Cu_2O_{10}$ (Ln = Ho, Y and Dy) were synthesized by HPHT method.

## 2. EXPERIMENTAL DETAILS

Samples of composition $RuSr_2(Ln_{3/4}Ce_{1/4})_2Cu_2O_{10}$ with Ln = Ho, Y and Dy were synthesised through a HPHT solid-state reaction route. For the HPHT synthesis and to fix the oxygen at 10.0 level, the molar ratio used were: $(RuO_2)$ + $(SrO_2)$ + $(SrCuO_2)$ + 3/4(CuO) + $1/4(CuO_{0.011})$ + $3/4(Ln_2O_3)$ + $1/2(CeO_2)$ resulting in $RuSr_2(Ln_{3/4}Ce_{1/4})_2Cu_2O_{10}$. $CuO_{0.011}$ is pure Cu-metal, for which precise oxygen content is determined before use. The materials were mixed in an agate mortar. Later around 300 mg of the mixture was sealed in a gold capsule and allowed to react in a flat-belt-type-high-pressure apparatus at 6GPa and 1200 $^{0}$C for 2 hours. Nearly no change was observed in the weight of synthesized samples, indicating towards their fixed nominal oxygen content. We believe the oxygen content of all the samples is close to nominal i.e. 10. Determination of the oxygen content of synthesized samples is yet warranted to know the oxygen value for these samples. X-ray powder diffraction patterns were obtained by a diffractometer (Philips-PW1800) with Cu $K_{\alpha}$ radiation. DC susceptibility data were collected by a SQUID magnetometer (Quantum Design, MPMS).

## 3. RESULTS AND DISCUSSION

$RuSr_2(Ln_{3/4}Ce_{1/4})_2Cu_2O_{10}$ with Ln = Ho, Y and Dy crystallised in a single-phase form in space group *I4/mmm* with lattice parameters $a = b = 3.819$ (1) Å, and $c = 28.439(1)$ Å for Ln = Y, $a = b = 3.813(2)$ Å, and $c = 28.419(1)$ Å for Ln = Ho, and $a = b = 3.824(4)$ Å, and $c = 28.445(1)$ Å for Ln = Dy. The volume of the cells is 413.2, 414.8 and 415.9 Å$^3$ for Ln = Ho, Y and Dy respectively. The trend of their cell volumes is in line with the rare earths ionic sizes. Figure 1 shows the X-ray diffraction patterns of finally synthesized Ln/Ru-1222 compounds. As seen from this figure these compounds are crystallised in a single phase form with only small amount of $SrRuO_3$ present in Ln = Y sample.

Figure 2 show both zero-field-cooled (zfc) and fc magnetic susceptibility versus temperature ($\chi$ vs. $T$) plots for the Y/Ru-1222 sample, in external fields of 5 and 20 Oe. As seen from this figure the fc magnetization curve shows an increase near 150 K, followed by a significant jump at around 100 K. The zfc branch shows a rise in magnetization at around 110 K and a cusp like down turn in magnetization at 100 K. In general the magnetization behaviour of the compound can be assigned to a weak ferromagnetic transition at around 100 K. However what is not understood is the initial rise of fc magnetization at 150 K. The interesting difference is that in HPHT synthesized Ln/Ru-1222 compounds the 150 K transition in fc magnetization is more pronounced than for reported Gd/Ru-1222 [1,12].

Figure 3 shows both zero-field-cooled (zfc) and fc magnetic susceptibility versus temperature ($\chi$ vs. $T$) plots for the Ln/Ru-1222 samples, with Ln = Ho, Dy. The general behaviour of the all the samples is similar to that as for Y/Ru-1222. The fc transition is seen in both the samples at 150 K. The zfc cusp and the diamagnetic transition are though Ln dependent, but essentially in the same temperature ranges.

Though the studied samples are almost single phase in x-ray, the minute impurities like $SrRuO_3$ or Ln/Ru-1212 might be responsible for the fc transition at 150 K. To exclude such a possibility we would like to stress that in Ln/Ru-1212 compounds the 150 K fc transition is followed by a cusp in zfc at same temperature and is also Ln dependent. For Ln = Ho and Dy in Ln/Ru-1212 the Ru spins magnetic ordering temperature is reported to be at 170 K [13]. In Ln/Ru-1222 compounds we do not observe a cusp in zfc and also the FC transition at 150 K is not Ln dependent. Hence the possible origin of fc transition at 150 K due to Ln/Ru-1212 is excluded. At this juncture we believe that the 150 K transition in fc magnetization of Ln/Ru-1222 compounds is intrinsic to this phase. This gets credence from the fact, that though Ln = Y sample unlike others contains small impurity of $SrRuO_3$, the 150 K transition in fc is same for all the studied compounds. In widely studied Gd/Ru-1222 compound also the rise in FC magnetization is reported at around 160-180 K, and was associated with an antiferromagnetic transition of Ru spins [1, 12].

We did couple of $M$ vs $H$ experiments for Ln = Ho sample at various temperatures of 130, 120, 110, 100 and 5 K, the results are shown in Fig.4 (a, b, c, d and e). At temperature of 130 and 120 K the $M$ vs. $H$ hysteresis loops exhibit an antiferromagnetic like structure with canted moments, though at 110 K the same possess more like an S-type spin-glass shape. It seems that the re-orientation of Ru-spins or change in canting angle takes place at 110 K. Further at 5 K, it is more like a ferromagnetic loop. The 5 K, data for non-magnetic Ln = Y will be discussed in Fig. 5. One wild speculation might be that before 100 K weak ferromagnetic transition, the Ru spins go through a spin-glass like transition at around 110 K and an antiferromagnetic transition at even higher temperature of say 150 K. Without detailed magnetic structure from neutron scattering experiments, it is difficult to comment on exact nature of the magnetism of various Ru-1222 compounds. Ironically, as we mentioned in the introduction, yet no detailed magnetic structure refinements from neutron scattering experiments are available for Ru-1222 compounds. Our current results are one step a head to widely reported Gd/Ru-1222 compound that the 150-160 K transition in magnetization before weak ferromagnetism at 110 K can not be left unnoticed as the same is quite sharp in our samples and is universal to all studied Ln/Ru-1222 compounds.

The zfc and fc significant branching temperature of 100 K for Y/Ru-1222 is relatively higher than previously reported ~ 80 K for Gd/Ru-1222. For reference, reported [8] $\chi$ vs. $T$ plot for Gd/Ru-1222 is shown in inset of Fig.2. Interestingly for magnetic ordering temperature for Gd/Ru-1212 of ~133 K was also found to be relatively lower than for HPHT synthesized

Y/Ru-1212 (~150 K) [4,5,13]. The zfc part of magnetic susceptibility at low temperature below 70 K shows a clear shoulder with further weak diamagnetic transition below ~50 K. The zfc curve did not show any diamagnetic transition ($T_d$) in $H$ = 100 Oe. The shoulder at 70 K is known as $T_c$ (superconducting transition temperature) from various experiments in Gd/Ru-1222. It is known earlier that due to internal magnetic field, these compounds are in a spontaneous vortex phase (SVP) even in zero external field [14]. For $T_d < T < T_c$ the compound remains in mixed state. Hence though superconductivity is achieved at relatively higher temperatures the diamagnetic response is seen at much lower $T$ and that also in quite small applied magnetic fields [14,15].

The isothermal magnetization as a function of magnetic field at 5 K with higher applied fields; 70000 Oe $\leq H \leq$ 70000 Oe for Ln = Y sample is shown in Fig.5. The saturation of the isothermal moment appears to occur above say 4 T applied fields. The presence of the ferromagnetic component is confirmed by hysteresis loops being observed at 5 K in the $M$ vs. $H$ plots, (see inset Fig. 2). Ru spins order magnetically above say 100 K with a ferromagnetic component within ($M_{rem}$, $H_c$ = 0.30 $\mu_B$, 150 Oe) at 5 K. As far the value of higher field (> 4 T) saturation moment is concerned, one can not without ambiguity extract the value for Ru contribution, because the contribution from Cu can not be ignored. In an under-doped HTSC compound Cu contributes an unknown paramagnetic signal to the system. Without considering the Cu contribution an effective moment of ~ 0.80 $\mu_B$ is obtained for Ru. This value is lees than for $Ru^{5+}$ low spin state ordering. In Y/Ru-1212 compound, based on various magnetization data effective moment of nearly 1$\mu_B$ is observed [5, 16].

**4. CONCLUSION**

In summary, single phase compounds of composition $RuSr_2(Ln_{3/4}Ce_{1/4})_2Cu_2O_{10}$ with Ln = Ho, Y and Dy were synthesised successfully through a HPHT solid-state reaction route for first time and their brief magnetization data is reported and discussed. Ln/Ru-1222 compounds being reported here with non magnetic Ln like Y could help in batter understanding of the magnetic structure of these compounds. Neutron scattering experiments along with various other physical property experiments on such a sample are warranted.

## 5. ACKNOWLEDGEMENT

Authors would like to thank I. Felner from Israel for his valuable suggestions and careful reading of the manuscript. VPS Awana at NIMS is supported by NIMS Fellowship.

## FIGURE CAPTIONS

Figure 1. Room temperature X-ray diffraction pattern for Ln/Ru-1222 compounds.

Figure 2. Magnetic susceptibility versus temperature ($\chi$ vs. $T$) plots for Y/Ru-1222 sample, in various applied fields of 5, and 20 Oe, inset shows the same for reported Gd/Ru-1222 compound.

Figure 3. Magnetic susceptibility versus temperature ($\chi$ vs. $T$) plots for (a) Dy/Ru-1222 and (b) Ho/Ru-1222, samples.

Figure 4. $M$ vs. $H$ plot for the Ho/Ru-1222 compound at $T$ = (a) 130 K, (b) 120 K, (c) 110 K, (d) 100 K and (e) 5 K.

Figure 5. $M$ vs. $H$ plot for the Y/Ru-1222 compound at $T$ = 5 K, the applied field H are in the range of -70 kOe < $H$ < 70 kOe. The inset shows the $M$ vs. $H$ plot for the same at 5 K, with -50000 ≤ H ≤ 5000 Oe.

*Figure 1. V.P.S. Awana and E. Takayama-Muromachi*

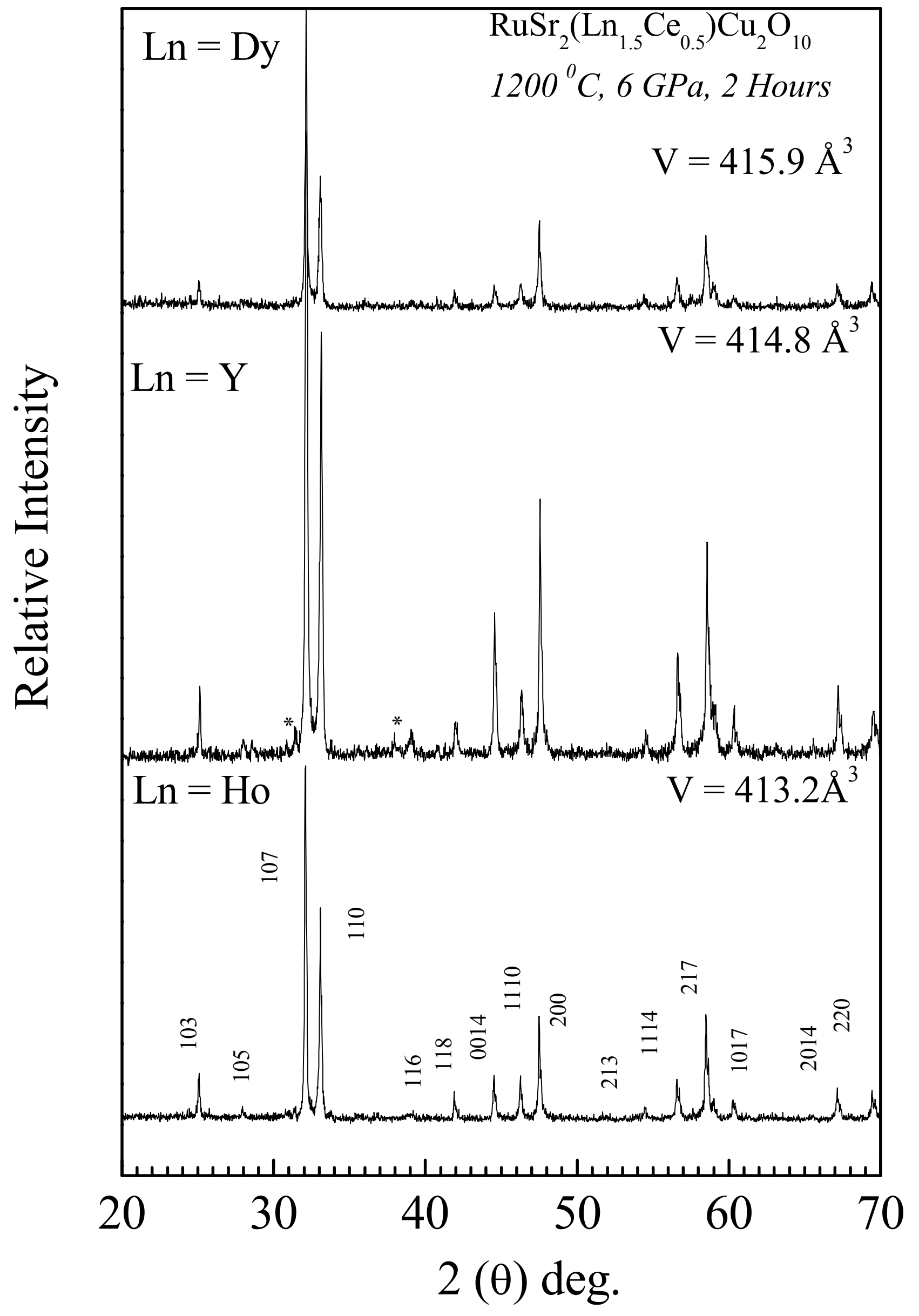

*Figure 2. V.P.S. Awana and E. Takayama-Muromachi*

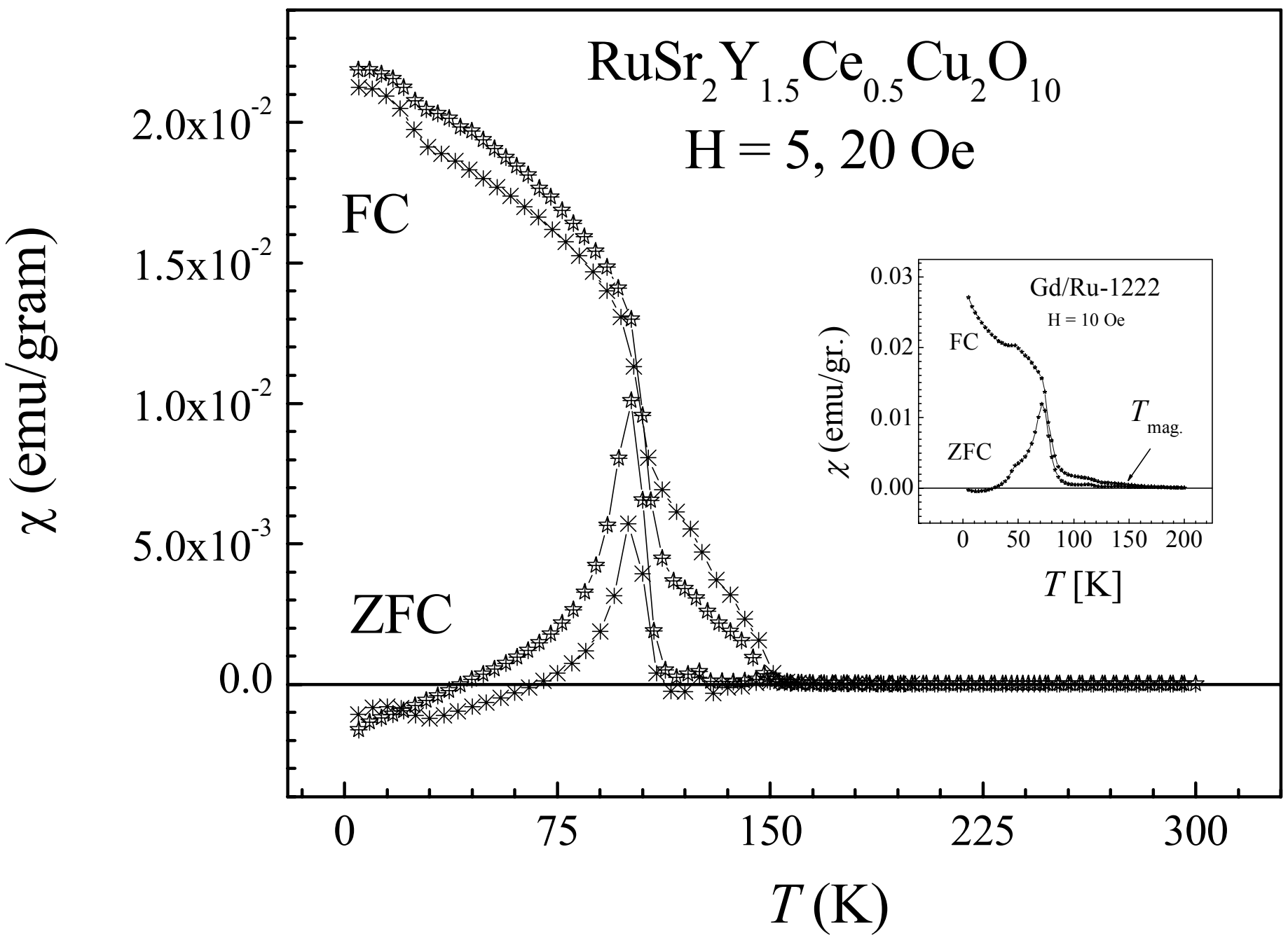

*Figure 3 (a). V.P.S. Awana and E. Takayama-Muromachi*

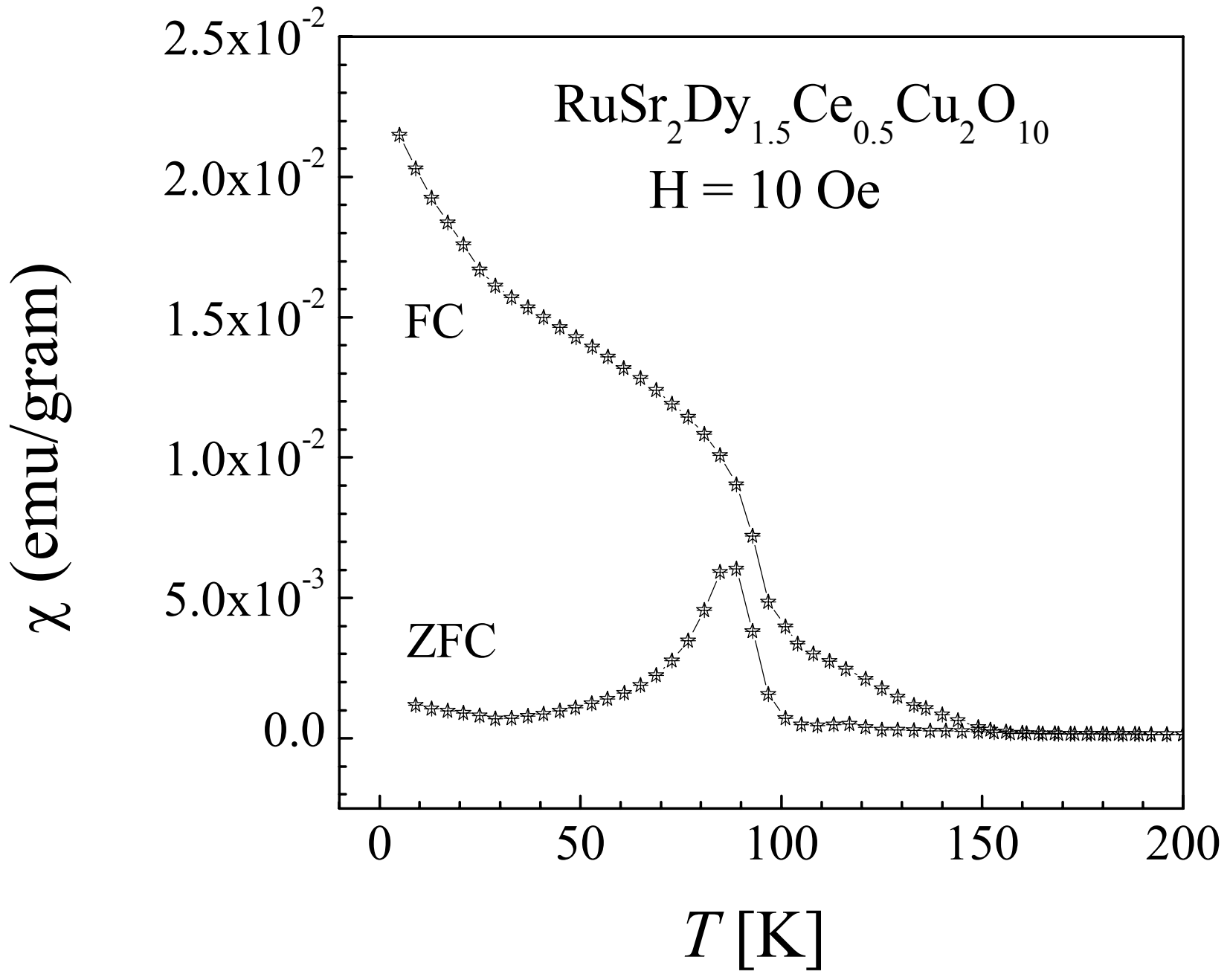

*Fig. 3 (b). V.P.S. Awana and E. Takayama-Muromachi*

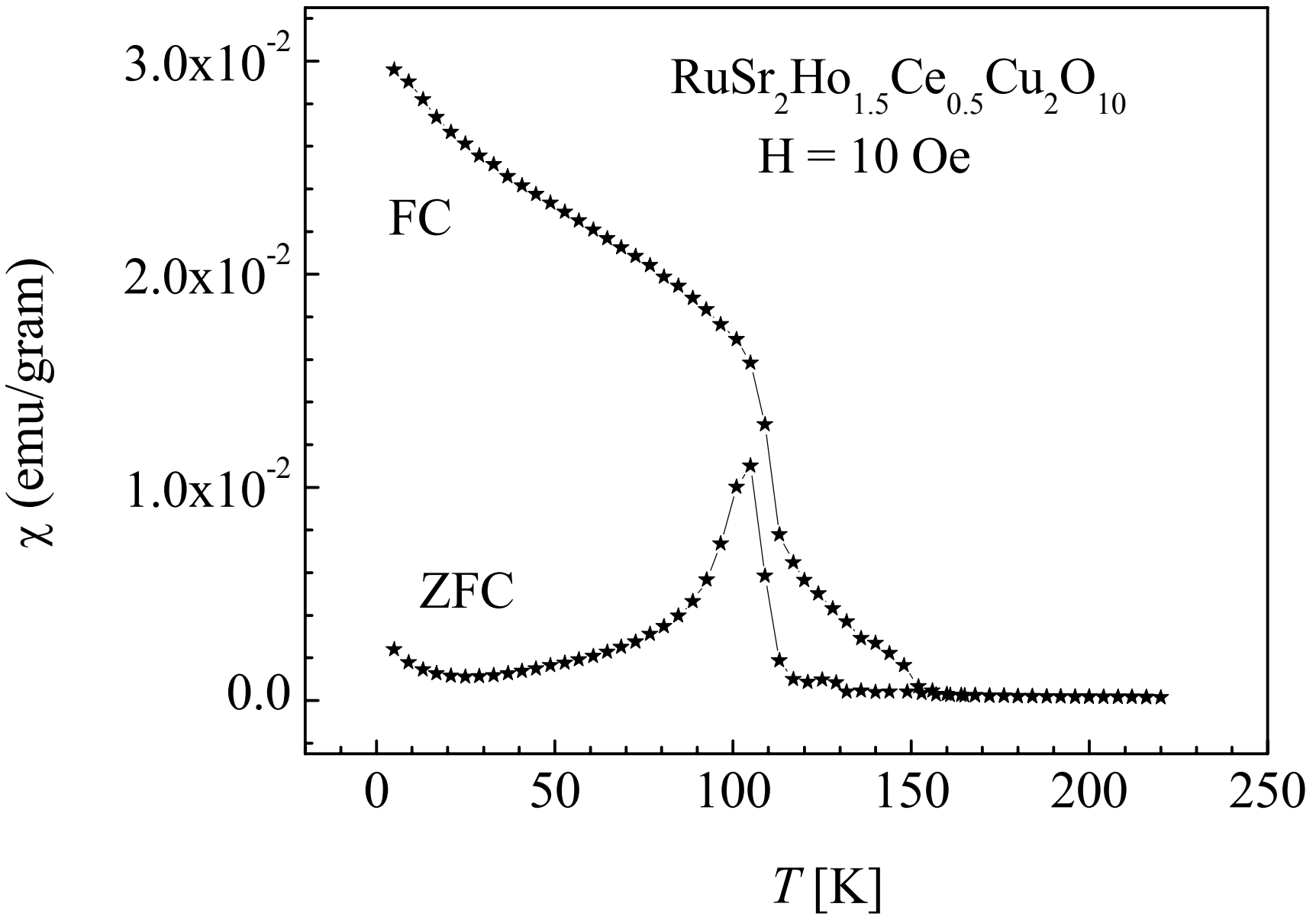

*Figure 4 (a, b, c, d and e). V.P.S. Awana and E. Takayama-Muromachi*

*(a)*

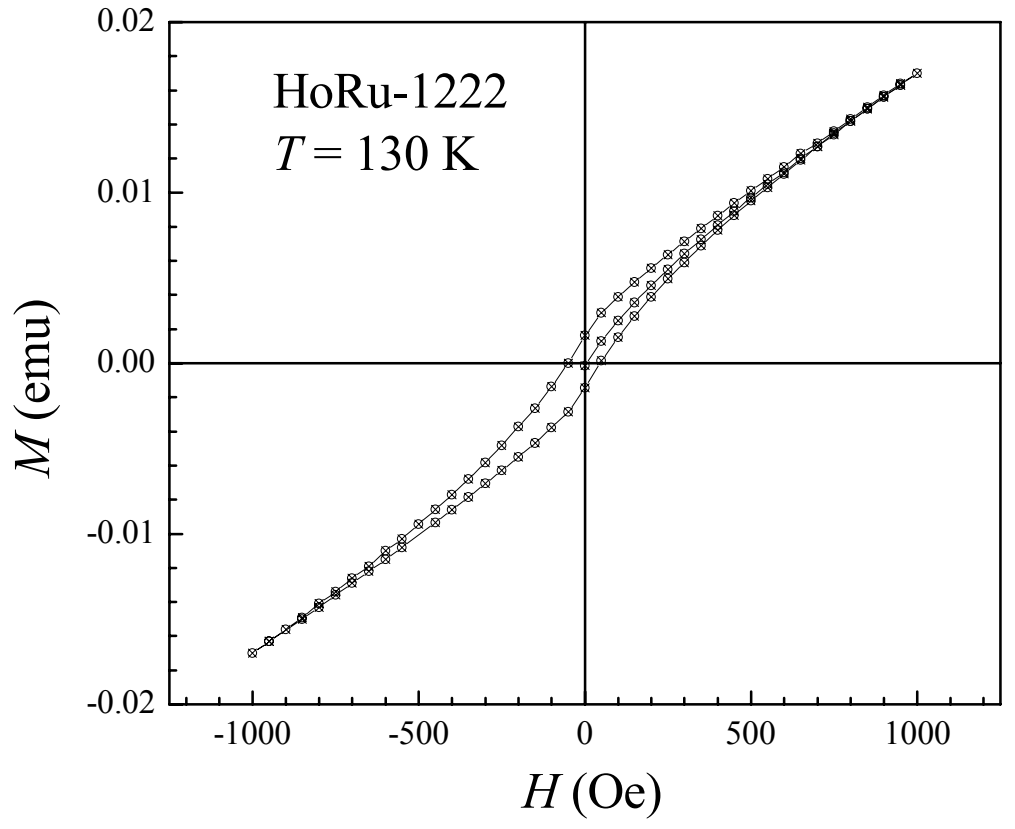

*(b)*

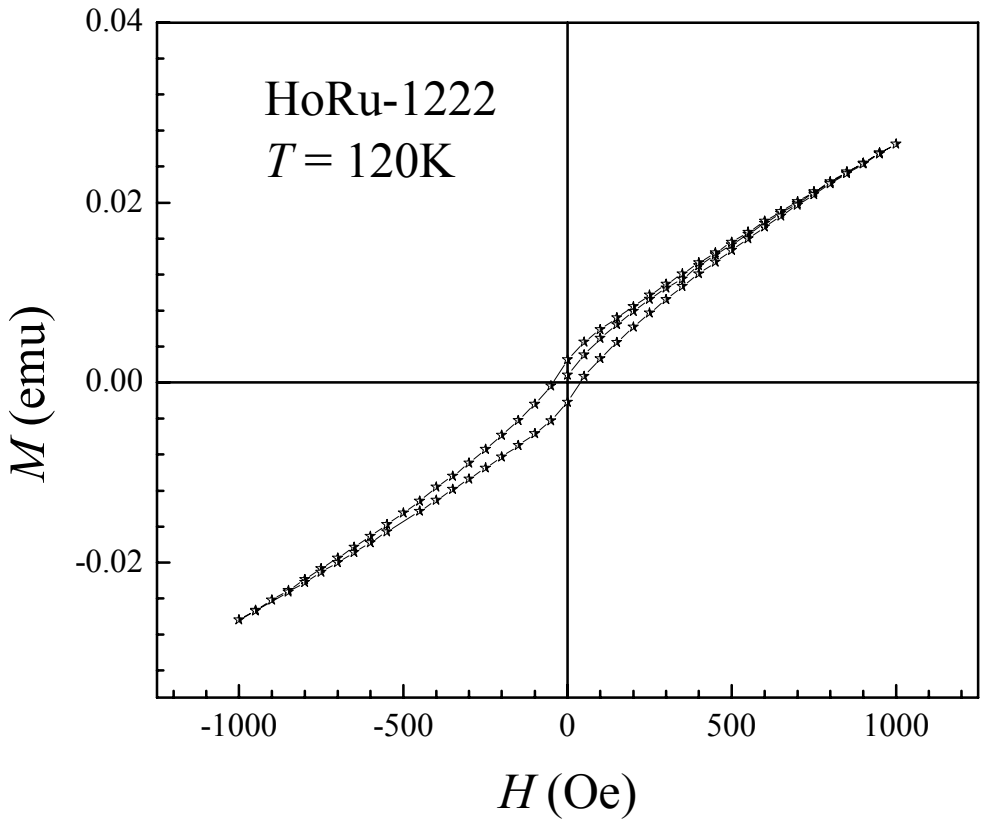

*(c)*

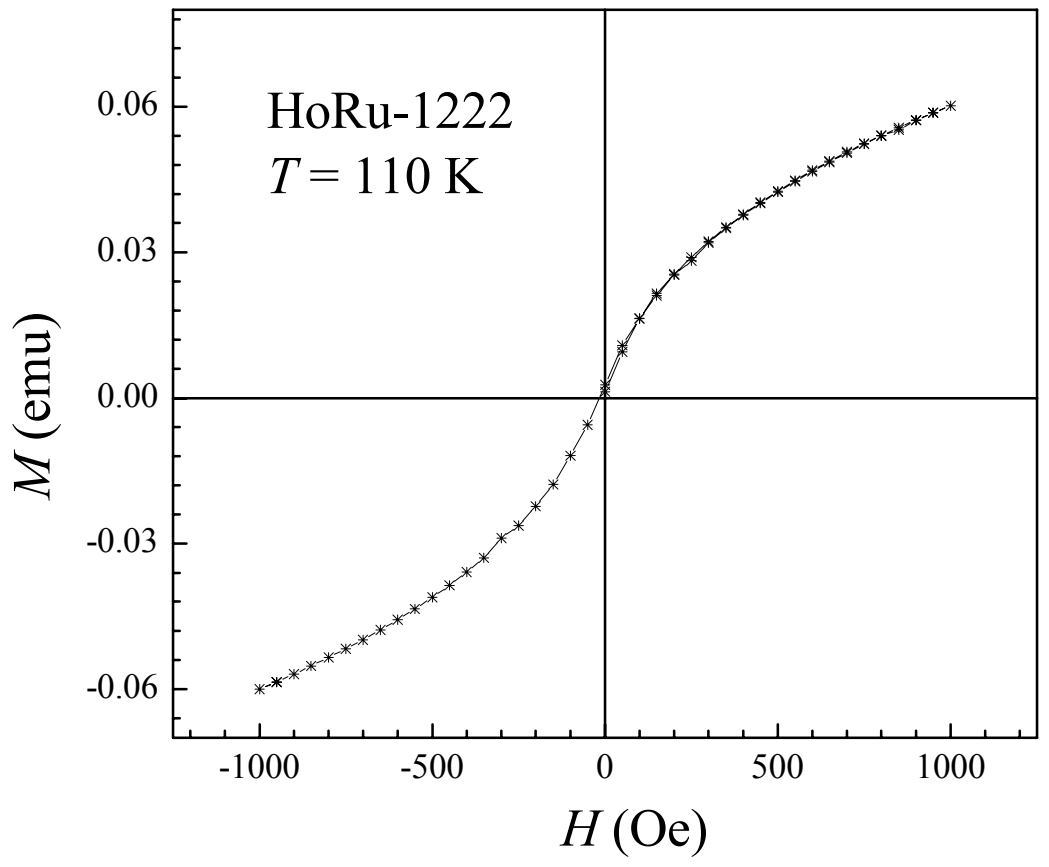

*(d)*

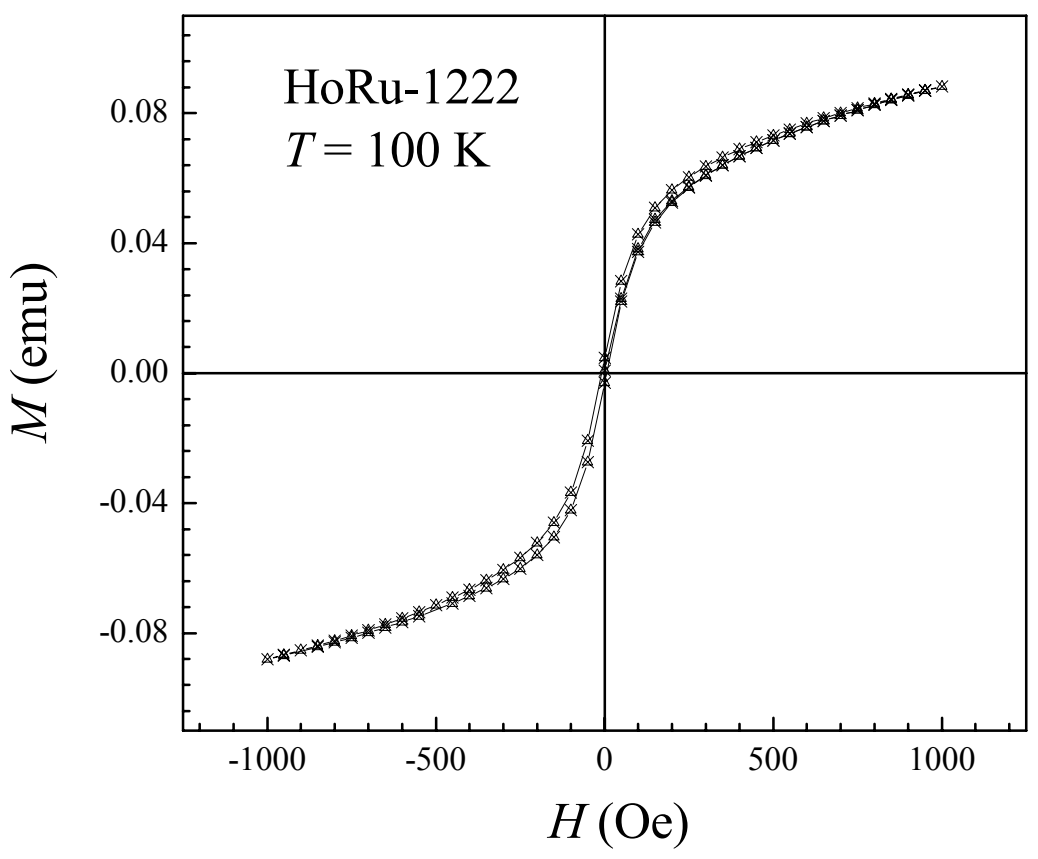

*(e)*

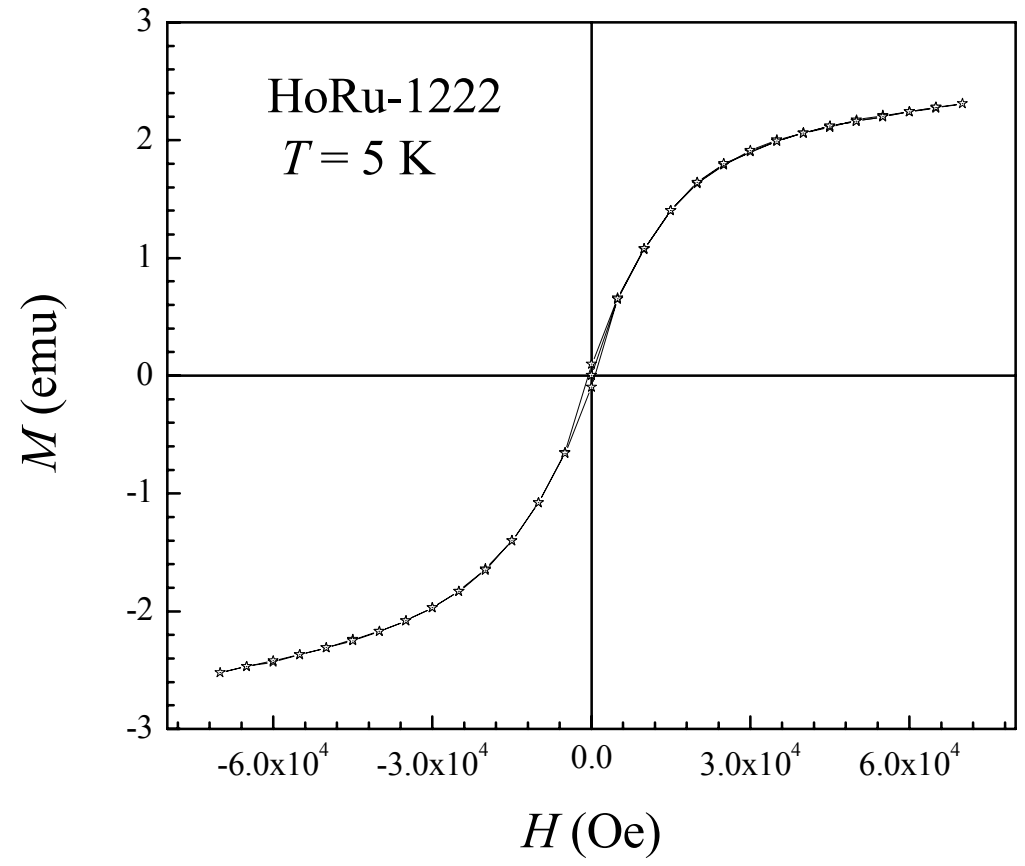

*Figure 5. V.P.S. Awana and E. Takayama-Muromachi*

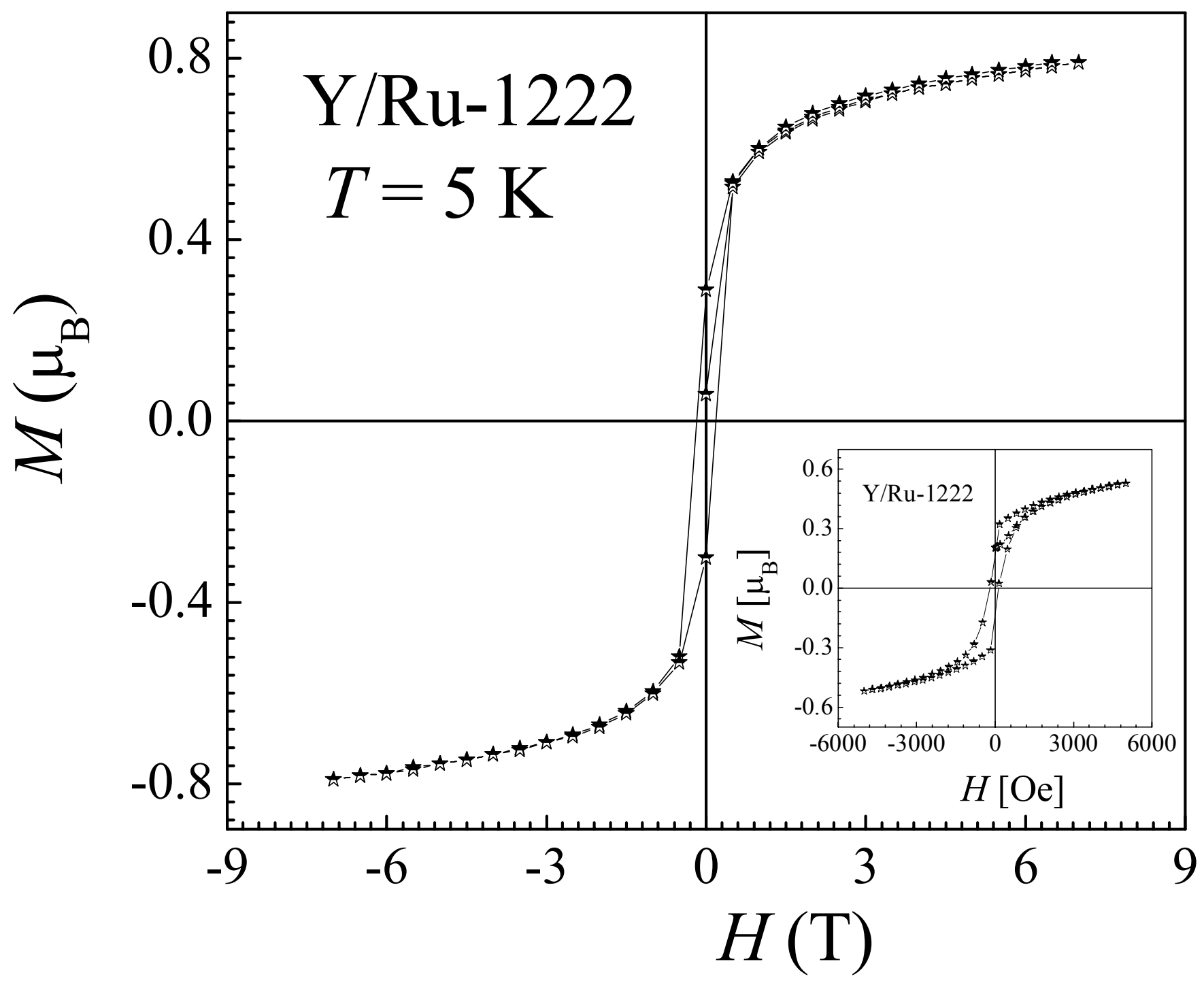